%% file: main.tex
\documentclass[sigconf,nonacm]{acmart}
\AtBeginDocument{%
  }

\usepackage{enumitem}
\usepackage[table]{xcolor}

\begin{document}

\title{PathoMIC: A Benchmark for Cross-Species Antimicrobial Peptide Activity Prediction}


\author{Yeqing Lu}
\affiliation{%
  \institution{University of Science and Technology of China}
  \state{Anhui}
  \city{Hefei}
  \country{China}
}
\email{luyeqing21@mail.ustc.edu.cn}

\author{Xiaoyan Zhao}
\affiliation{%
  \institution{Yong Loo Lin School of Medicine,\\National University of Singapore}
  \country{Singapore, Singapore}
}
\email{xyzhao@nus.edu.sg}

\author{Fuli Feng}
\affiliation{%
  \institution{University of Science and Technology of China}
  \state{Anhui}
  \city{Hefei}
  \country{China}
}
\email{fulifeng93@gmail.com}

\renewcommand{\shortauthors}{Lu and Feng}

\input{sections/abstract.tex}





\maketitle

\input{sections/intro.tex}

\input{sections/related_work.tex}

\input{sections/Dataset_Construction.tex}

\input{sections/Methods.tex}

\input{sections/Evaluation_Framework.tex}

\input{sections/Experiments.tex}

\input{sections/Conclusion_and_Future_Work.tex}

\bibliographystyle{ACM-Reference-Format}
\bibliography{reference}

\appendix

\section{Analysis of Pathogen Embeddings}

We examine whether the pathogen representations themselves encode
taxonomic relatedness and how this structure changes after the species adapter
or taxonomy GNN. 
We compare four checkpoint-aligned representations: (i) the frozen
768-dimensional PubMedBERT description embedding; (ii) the 128-dimensional
output of the trained SP species-adapter immediately before peptide--pathogen
fusion; (iii) the 640-dimensional hierarchy-fused output of the trained
taxonomy GNN; and (iv) the corresponding 640-dimensional GNN-RI output,
whose random input node vectors are sampled once and kept frozen. There are 366 pathogens 
in our filtered dataset. We therefore report representations of these pathogens embedding.

For visualization, we first $\ell_2$-normalize each representation and reduce
it to $\min(50,N-1,d)$ dimensions using PCA, where $N$ is the number of
pathogens and $d$ is the representation dimension. We then independently fit a
two-dimensional t-SNE projection to each PCA representation using perplexity
30, PCA initialization, automatic learning-rate selection, 2,000 iterations,
and random seed 42. Identical settings are used for all models.

To verify that apparent clusters are not solely projection artifacts, we also
measure genus $k$-nearest-neighbor purity in the original normalized
representation spaces using cosine distance. For a query pathogen, purity is
the fraction of its $k$ nearest pathogens that belong to the same genus; we
average this quantity over the 282 pathogens that have siblings in the same genus. The
frequency-derived random expectation is 2.02\%, obtained by averaging
$(n_g-1)/(N-1)$ over these queries, where $n_g$ is the number of pathogens in
the query genus. 
We estimate the 95\% confidence interval for 1-NN purity by
resampling pathogens with replacement 10,000 times and taking the 2.5th and
97.5th percentiles.

\begin{figure*}[t]
  \centering
  \begin{minipage}[t]{0.49\textwidth}
    \vspace{0pt}
    \centering
    \includegraphics[width=\linewidth]{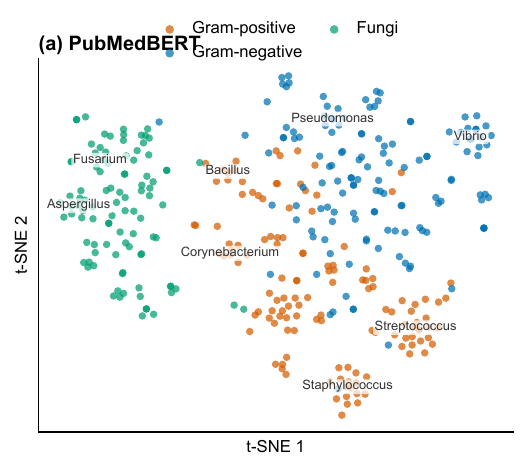}
  \end{minipage}\hfill
  \begin{minipage}[t]{0.49\textwidth}
    \vspace{0pt}
    \centering
    \includegraphics[width=\linewidth]{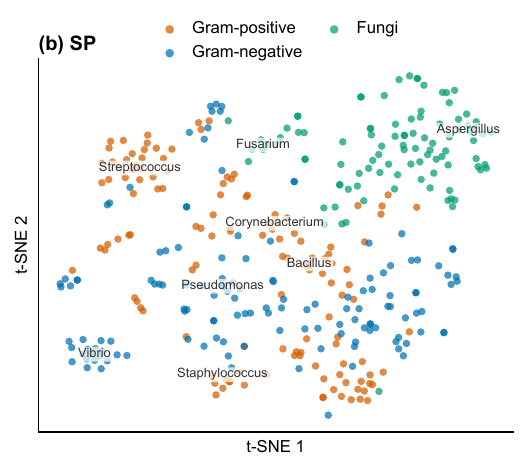}
  \end{minipage}

  \begin{minipage}[t]{0.49\textwidth}
    \vspace{0pt}
    \centering
    \includegraphics[width=\linewidth]{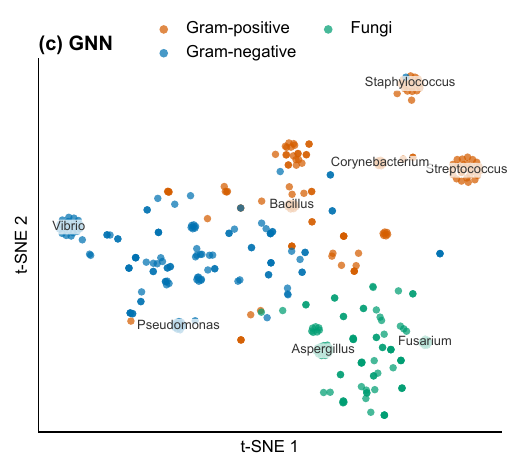}
  \end{minipage}\hfill
  \begin{minipage}[t]{0.49\textwidth}
    \vspace{0pt}
    \centering
    \includegraphics[width=\linewidth]{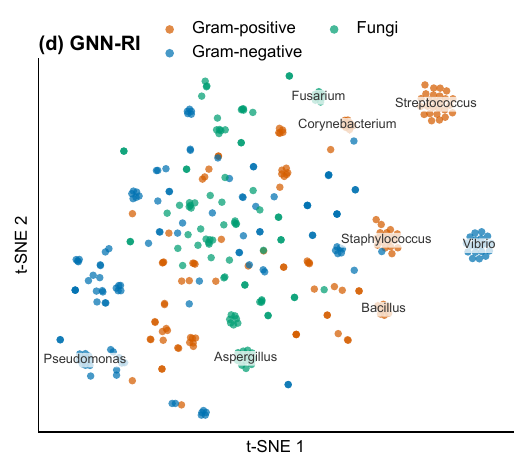}
  \end{minipage}
  \Description{Four t-SNE scatter plots compare raw PubMedBERT, SP-adapter,
  trained-GNN, and GNN-RandomInit representations for the same 366 pathogens.
  Colors distinguish Gram-positive bacteria, Gram-negative bacteria, and fungi;
  selected frequent genera are labeled.}
  \caption{Pathogen representation spaces for the shared training/inference
  cohort over 366 pathogens. (a) Frozen 768-dimensional PubMedBERT embeddings. (b) The
  128-dimensional SP species-adapter outputs. (c) The 640-dimensional
  hierarchy-fused outputs of the trained taxonomy GNN. (d) The corresponding
  GNN-RI outputs obtained from frozen random node features. All panels
  contain the same 366 pathogens and use $\ell_2$ normalization, PCA-50, and
  identical t-SNE hyperparameters. Colors encode pathogen group and labels
  mark selected frequent genera.}
  \label{fig:embedding-taxonomy}
\end{figure*}

Figure~\ref{fig:embedding-taxonomy}(a) shows that the raw PubMedBERT space is
already taxonomically organized. Fungi occupy a region largely separated from
the bacterial pathogens, while several frequent genera form visually compact
groups, including \textit{Streptococcus}, \textit{Staphylococcus},
\textit{Vibrio}, \textit{Aspergillus}, and \textit{Fusarium}. This organization
is substantially stronger than expected by chance in the original 768-D
space: PubMedBERT achieves 92.20\% genus purity at $k=1$, compared with the 2.02\% random expectation. 
Equivalently,
260 of the 282 evaluated pathogens have a nearest neighbor from the same
genus. Purity decreases as $k$ grows because many genera contain fewer than ten
represented pathogens, forcing larger neighborhoods to include other genera.
Together, the visual and high-dimensional results demonstrate that the fixed
PubMedBERT descriptions make close taxonomic relatives readily recoverable
without using the taxonomy graph.

The SP adapter preserves much of this organization but also visibly reshapes
the space (Figure~\ref{fig:embedding-taxonomy}(b)). Its genus purities decrease
to 79.79\% at $k=1$, respectively. Thus, the
adapter does not improve MIC prediction merely by making same-genus clusters
tighter. Instead, its task-specific transformation appears to trade some
taxonomic neighborhood structure for other description-derived,
species-specific signals that are useful to the regression objective.
The trained GNN produces the clearest genus-level separation
as shown in Figure~\ref{fig:embedding-taxonomy}(c), reaching 99.65\% purity at $k=1$. 
However, GNN-RI exhibits a very similar
structure shown in Figure~\ref{fig:embedding-taxonomy}(d) and obtains 97.87\% purity at $k=1$.
This result demonstrates the GNN explicitly recon-
structs and strengthens this organization through graph topology.

\section{Complete Pathogen Taxonomy}

\begin{figure*}[p]
  \centering
  \includegraphics[width=0.78\textwidth]{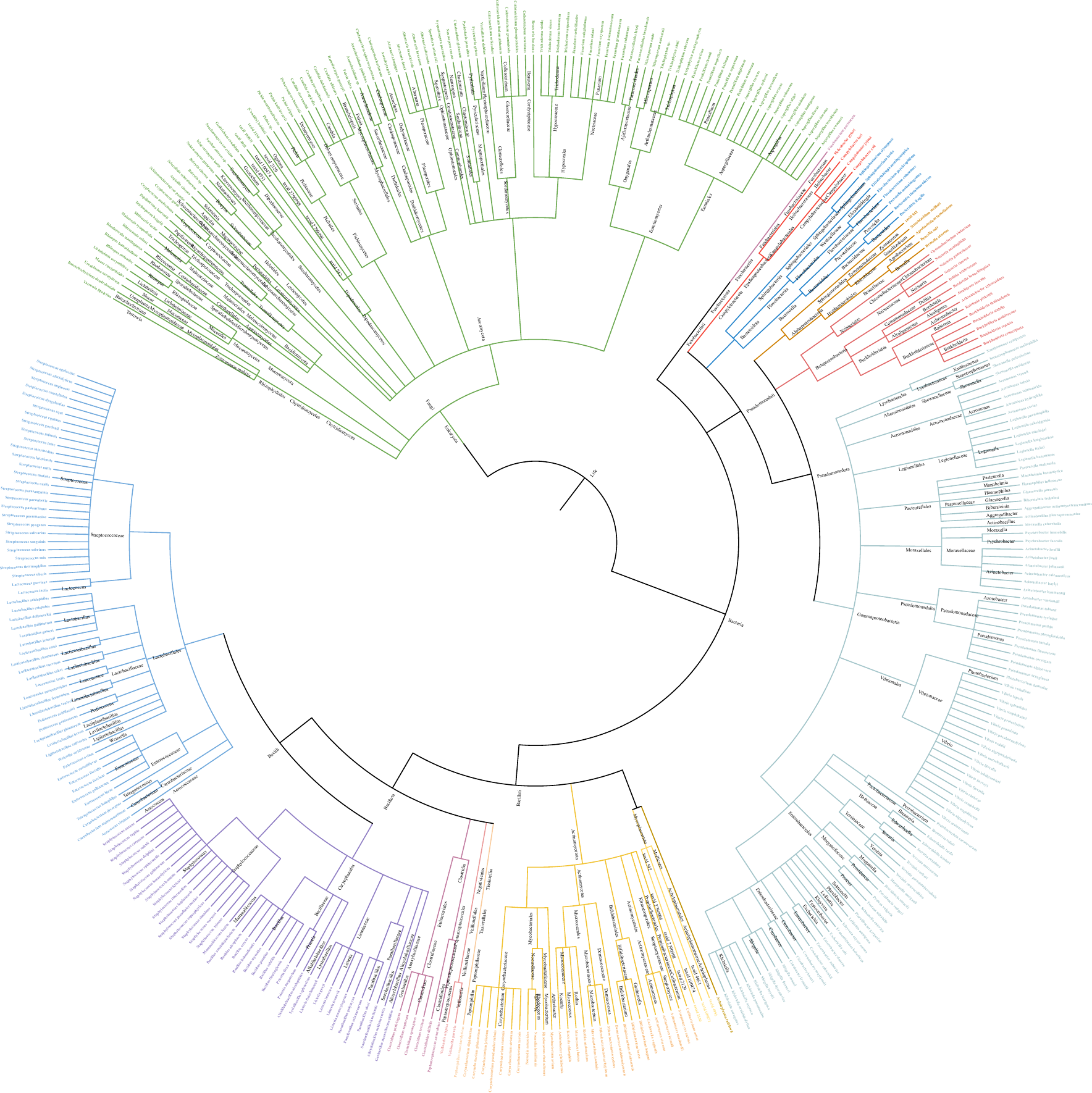}
  \Description{A circular taxonomy tree showing the hierarchical relationships
  among all pathogens included in PathoMIC.}
  \caption{Overview of the complete pathogen taxonomy represented in PathoMIC.
  Branches encode hierarchical taxonomic relationships; enlarged sectors are
  provided in Figures~\ref{fig:taxonomy-sector-ul}--%
  \ref{fig:taxonomy-sector-lr}.}
  \label{fig:complete-taxonomy}
\end{figure*}

Figure~\ref{fig:complete-taxonomy} provides an overview of the complete
taxonomy graph in PathoMIC. We report all 366 pathogen species included in filtered 
dataset used in the benchmark setting for model comparison. Because the leaf labels are necessarily
dense at this scale, Figures~\ref{fig:taxonomy-sector-ul}--%
\ref{fig:taxonomy-sector-lr} enlarge four overlapping sectors of the same
vector graphic. The overlap preserves local context around sector boundaries.

\begin{figure*}[p]
  \centering
  \includegraphics[width=0.92\textwidth,viewport=0 400 448 849,clip]
    {figs/taxomony_graph.pdf}
  \Description{An enlarged view of the upper-left sector of the complete
  circular pathogen taxonomy.}
  \caption{Enlarged upper-left sector of the PathoMIC pathogen taxonomy.}
  \label{fig:taxonomy-sector-ul}
\end{figure*}

\begin{figure*}[p]
  \centering
  \includegraphics[width=0.92\textwidth,viewport=400 400 849 849,clip]
    {figs/taxomony_graph.pdf}
  \Description{An enlarged view of the upper-right sector of the complete
  circular pathogen taxonomy.}
  \caption{Enlarged upper-right sector of the PathoMIC pathogen taxonomy.}
  \label{fig:taxonomy-sector-ur}
\end{figure*}

\begin{figure*}[p]
  \centering
  \includegraphics[width=0.92\textwidth,viewport=0 0 448 448,clip]
    {figs/taxomony_graph.pdf}
  \Description{An enlarged view of the lower-left sector of the complete
  circular pathogen taxonomy.}
  \caption{Enlarged lower-left sector of the PathoMIC pathogen taxonomy.}
  \label{fig:taxonomy-sector-ll}
\end{figure*}

\begin{figure*}[p]
  \centering
  \includegraphics[width=0.92\textwidth,viewport=400 0 849 448,clip]
    {figs/taxomony_graph.pdf}
  \Description{An enlarged view of the lower-right sector of the complete
  circular pathogen taxonomy.}
  \caption{Enlarged lower-right sector of the PathoMIC pathogen taxonomy.}
  \label{fig:taxonomy-sector-lr}
\end{figure*}

\end{document}

%% file: sections/abstract.tex
\begin{abstract}
With activity against multidrug-resistant pathogens and mechanisms distinct from conventional antibiotics, antimicrobial peptides (AMPs) offer a promising approach to combating antibiotic-resistant infections.
However, their potency varies substantially across pathogen species, making accurate prediction of the minimum inhibitory concentration (MIC) for specific peptide–pathogen pairs essential for prioritizing candidates before costly experimental validation. Existing predictors are trained mainly on a few well-represented pathogens and rarely exploit biological relationships across species, limiting their generalization to low-resource and unseen pathogens. We introduce PathoMIC, the largest and most pathogen-diverse unified dataset for quantitative antimicrobial peptide activity prediction, containing 74,751 experimentally reported MIC measurements across 424 pathogen species. PathoMIC integrates peptide sequences, standardized MIC values, pathogen descriptions, and taxonomic relationships to facilitate knowledge transfer across related species. We establish few-shot and zero-shot cross-species evaluation protocols and develop a knowledge-enhanced framework that leverages pathogen descriptions and taxonomy. The framework yields substantial improvements for low-resource species with limited supervision, while gains for entirely unseen species remain modest, highlighting the difficulty of zero-shot cross-species MIC prediction. PathoMIC provides a standardized foundation for cross-species activity modeling and pathogen-specific virtual screening. Code is available at \url{https://anonymous.4open.science/r/PathoMIC-546D/}.
\begin{figure}[t]
  \centering
  \includegraphics[width=\columnwidth]{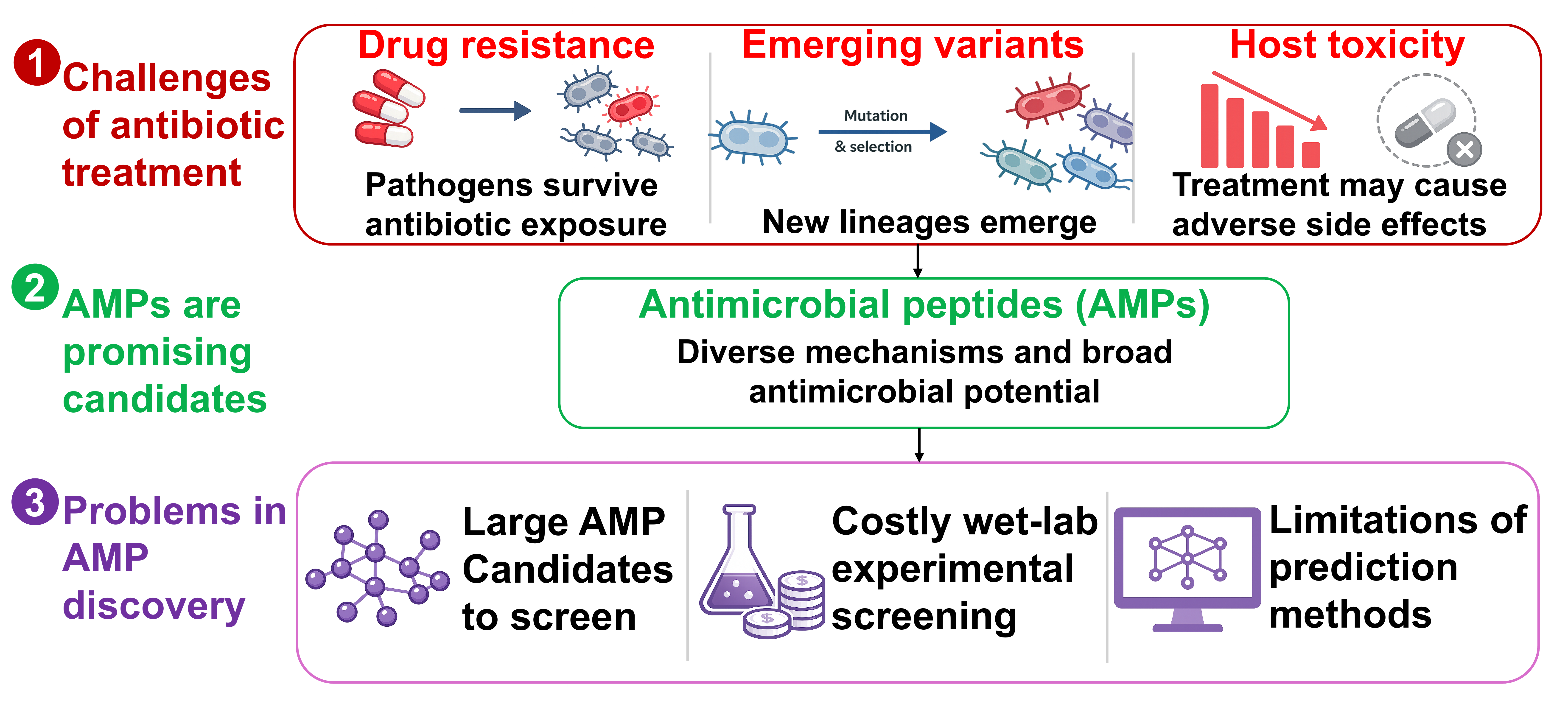}
  \caption{Challenges of conventional antibiotic treatment and problems in Antimicrobial peptide discovery.}
  \Description{Three panels illustrate drug-resistant pathogens, emerging pathogen variants, and adverse effects of antibiotic treatment on the human host. Arrows connect these challenges to the need for new antimicrobial candidates and to antimicrobial peptides.}
  \label{fig:antibiotic_challenges}
\end{figure}
\end{abstract}

%% file: sections/intro.tex
\section{Introduction}

Pathogenic microorganisms pose a continuously evolving threat to human health. Through spontaneous mutations, horizontal gene transfer, and selection under antimicrobial exposure, pathogens can acquire adaptive traits and resistance mechanisms, giving rise to emerging variants and drug-resistant lineages that are increasingly difficult to control~\cite{WHO,Holmes2016AMR}. Conventional antibiotic treatments also pose risks to patients. Broad-spectrum antibiotics may cause adverse drug reactions and organ toxicity, disrupt commensal microbial communities, reduce colonization resistance, and increase susceptibility to opportunistic or recurrent infections~\cite{2023Antibiotic,Blaser2016Microbiome}. Antimicrobial peptides (AMPs), which exhibit diverse mechanisms of action and broad-spectrum activity, have therefore emerged as promising alternatives to conventional antibiotics~\cite{2020Antimicrobial,Mookherjee2020HDP}.

The rapid accumulation of experimentally characterized AMP sequences, together with advances in artificial intelligence, has stimulated extensive research on computational AMP discovery and design~\cite{DBAASP,DRAMP}. Existing studies have explored vast peptide sequence spaces and generated novel AMP candidates~\cite{grampa,2025HMAMP,AMPGAN,HydrAMP,2025DLFea4AMPGen}. However, candidate generation alone is insufficient: practical AMP development also requires reliable computational tools for identifying peptides with strong activity against specific pathogens~\cite{AMP-Designer}. The minimum inhibitory concentration (MIC), defined as the lowest concentration of an antimicrobial agent that prevents visible microbial growth under specified experimental conditions, is one of the most widely used quantitative measures of antimicrobial activity~\cite{Andrews2001MIC}. Accurate MIC prediction can therefore substantially improve computer-assisted AMP development by prioritizing promising peptide--pathogen pairs before costly experimental validation~\cite{Chung2024MIC}.

Existing computational screening methods fall into two broad categories. Early approaches treat AMP screening as a classification task, predicting whether a peptide is antimicrobial or assigning it to a discrete activity level~\cite{2021Alignment,2023Gut}. More recent approaches directly regress continuous MIC values from peptide sequences~\cite{grampa,LSTM_Seq,2025HMAMP,BERTAmPEP60}. Despite this progress, most existing MIC predictors are developed and evaluated on only one or a few well-represented pathogens~\cite{grampa,2025HMAMP,BERTAmPEP60,Chung2024MIC,Chiu2026ANIA}, resulting in two fundamental limitations. First, current MIC measurements cover only a limited range of pathogens with abundant observations, while most low-resource pathogens remain underrepresented. Second, conventional MIC predictors primarily model antimicrobial potency from peptide sequences and manually engineered features, often by training separate models for individual pathogens. By treating pathogen identity merely as a categorical label, these approaches fail to exploit biological relationships among pathogens and cannot effectively transfer knowledge to low-resource or previously unseen species. Therefore, a comprehensive AMP--MIC dataset covering a broad spectrum of pathogens and a standardized benchmark for evaluating cross-species generalization are critically needed.

To address these limitations, we construct PathoMIC, a large-scale AMP--MIC dataset containing 74,751 MIC records across 424 pathogen species. The records are curated from multiple public AMP databases and the scientific literature, including APD6~\cite{APD}, DBAASP~\cite{DBAASP}, DRAMP~\cite{DRAMP}, CAMP3~\cite{CAMPR3}, YADAMP~\cite{YADAMP}, UniProt~\cite{UniProt}, HMAMP~\cite{2025HMAMP}, and GRAMPA~\cite{grampa}. Each record associates an AMP with its target pathogen and an experimentally reported MIC value. PathoMIC covers a broad spectrum of pathogens, including Gram-positive bacteria, Gram-negative bacteria, and fungi. Beyond categorical species identifiers, we collect textual descriptions of individual pathogens and retrieve their taxonomic relationships from the NCBI Taxonomy database~\cite{NCBI}. We organize these relationships into a structured taxonomy graph that explicitly represents evolutionary relatedness among pathogens. Together, the peptide sequences, quantitative activity measurements, pathogen descriptions, and taxonomic relationships form, to the best of our knowledge, the largest and most pathogen-diverse unified resource for studying quantitative AMP activity across pathogens.

Building on PathoMIC, we establish a standardized benchmark for cross-species MIC prediction under long-tailed supervision. Since available MIC measurements are concentrated on a small number of well-studied pathogens, practical AMP screening often requires transferring knowledge to species with limited or no labeled data. Conventional random record splits, however, largely test interpolation within observed pathogens and therefore do not adequately measure this capability. To better reflect real-world deployment, our benchmark includes few-shot prediction for low-resource pathogens and zero-shot prediction for species entirely unseen during training. We systematically evaluate representative MIC predictors and further propose a strong pathogen-aware framework that combines an ESM-2 peptide encoder~\cite{ESM2} with pathogen descriptions and taxonomic relationships. Extensive experiments show that the framework substantially improves prediction when limited target-species supervision is available, whereas improvements for entirely unseen pathogens are considerably smaller. These results demonstrate the utility of pathogen-specific knowledge in data-scarce settings while also exposing the remaining challenge of zero-shot cross-species transfer.
Our main contributions are summarized as follows:

\vspace{-5pt}
\begin{itemize}[leftmargin=*]
    \item We construct \textbf{PathoMIC}, the largest and most pathogen-diverse unified AMP--MIC resource, containing 74,751 records across 424 pathogen species and enriched with pathogen descriptions and structured taxonomic relationships for quantitative antimicrobial activity modeling.
    \item  We introduce a standardized benchmark for cross-species MIC prediction, with few-shot and zero-shot protocols that capture long-tailed pathogen distributions and directly assess generalization to low-resource and unseen species.
    \item We evaluate representative MIC prediction models under the proposed benchmark and propose a knowledge-enhanced framework that integrates peptide representations with pathogen descriptions and taxonomic information. The framework achieves state-of-the-art performance on PathoMIC and demonstrates the value of biological knowledge for cross-species MIC prediction.
\end{itemize}

%% file: sections/related_work.tex
\section{Related Work}

\subsection{MIC Prediction for AMP Screening}

Early computational approaches to AMP screening primarily formulated the task as coarse-grained classification rather than quantitative potency prediction. \citeauthor{2021Alignment}
developed alignment-free random forest models from physicochemical and
sequence-derived descriptors to distinguish AMPs from non-AMPs and to classify
antibacterial, antifungal, antiparasitic, and antiviral functional types
~\cite{2021Alignment}. Moving beyond binary classification, \citeauthor{2023Gut} categorized peptides as highly active, weakly active, or inactive based on experimentally reported MIC ranges. Their AutoGluon ensemble incorporated 94 handcrafted features,
including amino acid composition, hydrophobicity, amphiphilicity, and net charge,
to screen peptides against enteric pathogens~\cite{2023Gut}. Although effective for
candidate filtering, these classification formulations discretize antimicrobial
potency and cannot estimate continuous MIC values. Subsequent studies therefore formulated MIC prediction as a regression task. \citeauthor{grampa}
introduced GRAMPA and trained an ensemble of convolutional neural networks~\cite{CNN} operating on one-hot-encoded peptide sequences for joint AMP classification and MIC regression. Its quantitative predictor primarily modeled activity against \textit{Escherichia coli} and was subsequently coupled with simulated annealing to identify sequences with low predicted MIC values~\cite{grampa}. \citeauthor{LSTM_Seq} proposed a multi-stage screening pipeline consisting of
a classifier, a ranker, and an LSTM-based~\cite{lstm} MIC regressor. The regressor was further calibrated through incremental learning using internally measured peptides~\cite{LSTM_Seq}. With the emergence of protein language models~\cite{esm,2023Evolutionary,ProLLaMAAP,ProtGPT2,ProteinBERT}, recent methods have adopted pretrained protein 
representations for MIC prediction. HMAMP employed a Prot-BERT-BFD-based~\cite{ProteinBERT} MIC predictor trained on
\textit{E. coli} measurements as an activity objective for multi-objective peptide
design~\cite{2025HMAMP}. Similarly, BERT-AmPEP60 fine-tuned separate ProtBERT~\cite{ProteinBERT}
regressors for \textit{E. coli} and \textit{Staphylococcus aureus}
~\cite{BERTAmPEP60}.

\subsection{Pathogen-Aware AMP Data and Cross-Species Generalization}

\begin{figure*}[hbpt]
    \centering
    \includegraphics[width=0.98\textwidth,trim=0 135 0 0,clip]
        {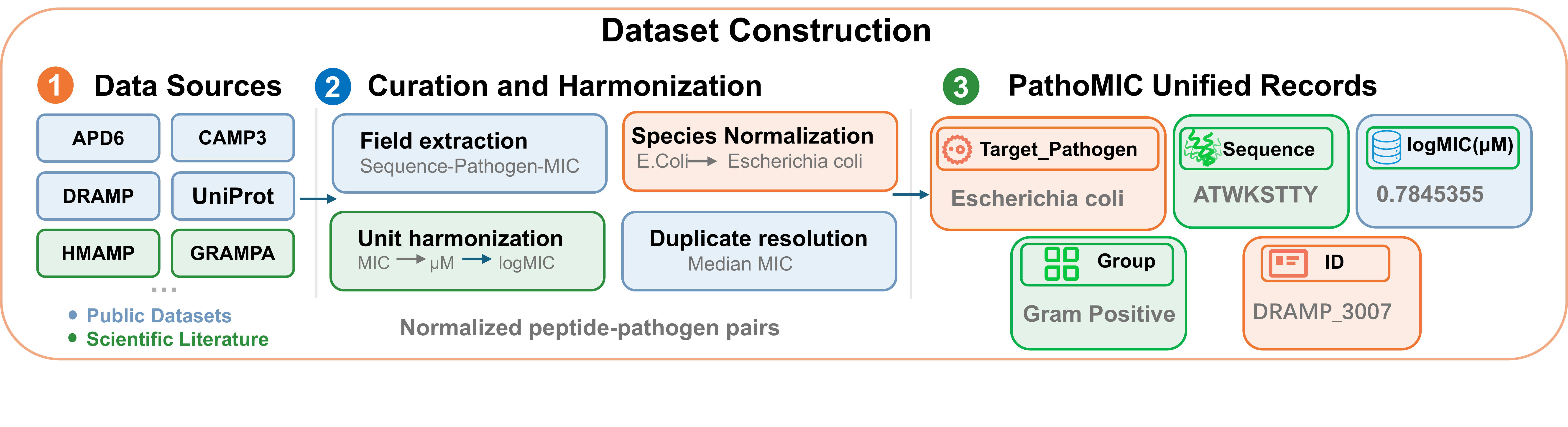}
    \smallskip

    {\small\textbf{(a)} PathoMIC data sources, curation and harmonization, and unified records.}

    \medskip
    \begin{minipage}[hbpt]{0.375\textwidth}
        \centering
        \includegraphics[width=\linewidth]{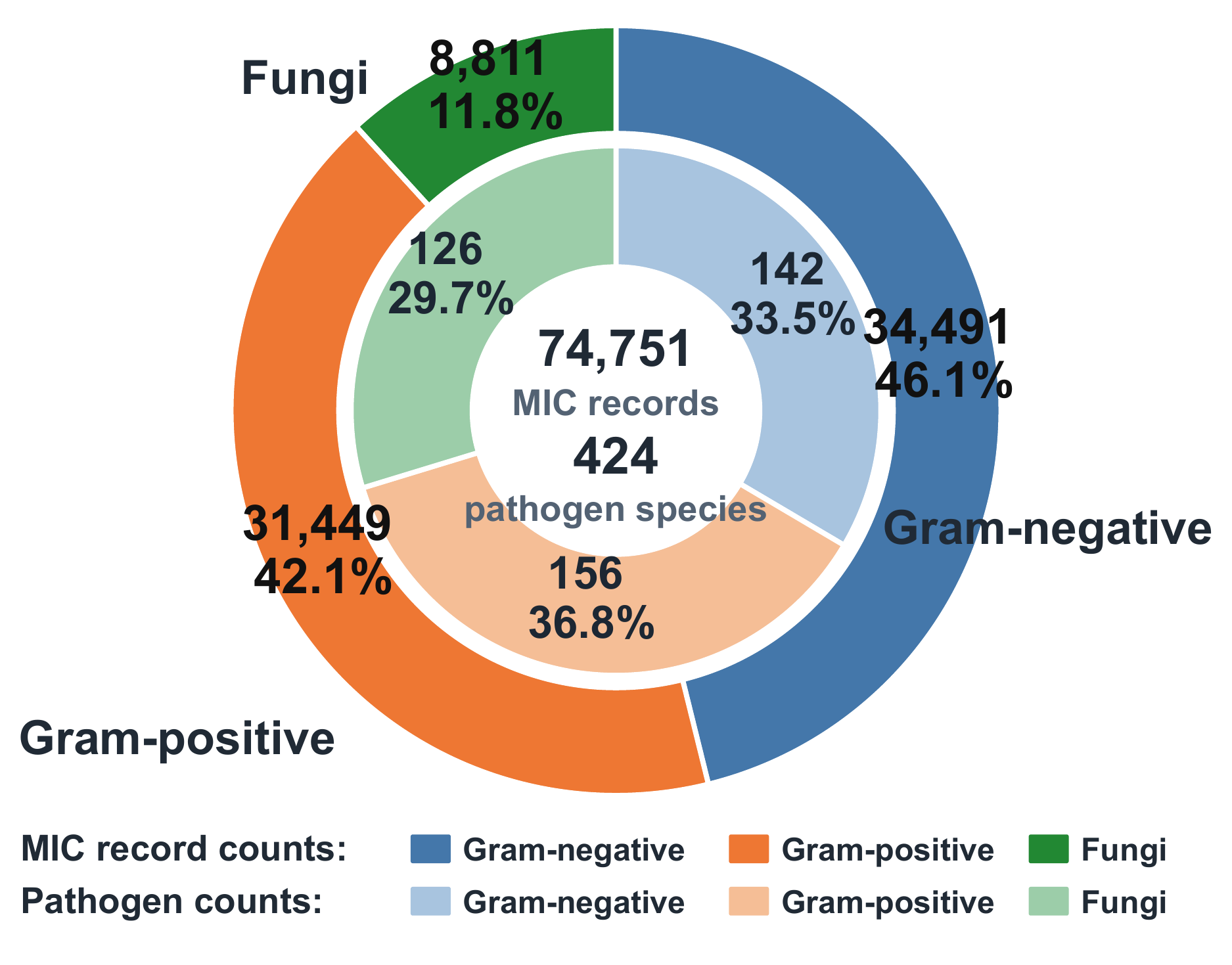}

        \medskip
        {\small\textbf{(b)} Composition of MIC records and pathogen species by pathogen group.}
    \end{minipage}
    \hfill
    \begin{minipage}[hbpt]{0.605\textwidth}
        \centering
        \includegraphics[width=\linewidth]{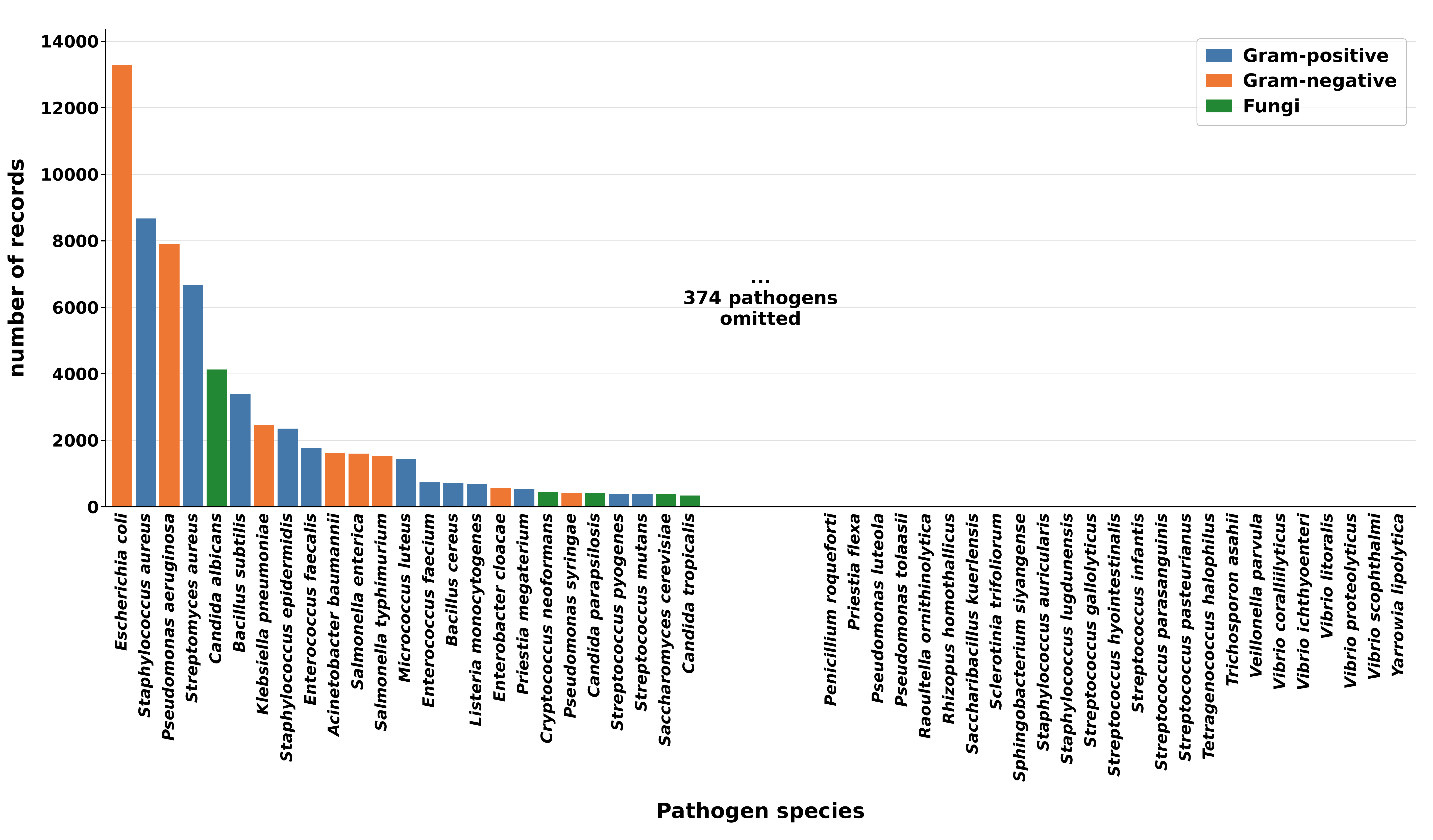}
        {\small\textbf{(c)} Long-tailed distribution of MIC records across pathogen species.}
    \end{minipage}

    \caption{Overview of PathoMIC construction and dataset statistics.}
    \Description{Panel (a) shows AMP databases and scientific literature being
    processed through field extraction, unit harmonization, species
    normalization, and duplicate resolution to produce unified PathoMIC records.
    Panel (b) summarizes MIC records and pathogen species across Gram-negative
    bacteria, Gram-positive bacteria, and fungi. Panel (c) shows the long-tailed
    distribution of MIC records across individual pathogen species.}
    \label{fig:dataset-curation-pipeline}
    \label{fig:dataset-composition}
\end{figure*}

A related pathogen-aware setting is microbial strain-specific AMP prediction. 
\citeauthor{Vishnepolsky2022} augmented peptide descriptors with target-strain
genome features and showed that incorporating organism-level information improves binary
activity prediction, including when examples from a test strain are excluded
from training~\cite{Vishnepolsky2022}. This result suggests
that relationships among microorganisms can facilitate knowledge transfer across strains and species. KPPepGen incorporates the knowledge of pathogens in the generation of conditional peptides
~\cite{KPPEPGEN}. However, it does not formulate quantitative MIC prediction and evaluation protocols. In contrast, our work consolidates peptide--pathogen MIC records with
substantially broader species coverage, augments every pathogen with textual and
taxonomic information, and defines standardized few--shot and zero--shot splits
for cross--species MIC regression. This benchmark enables controlled evaluation of whether predictive models can transfer quantitative antimicrobial knowledge
from well-represented pathogens to rare or previously unseen species.

%% file: sections/Dataset_Construction.tex
\section{Dataset Construction}

\subsection{Data Curation}
\begin{figure*}[htbp]
    \centering
    \begin{minipage}[t]{0.49\textwidth}
        \centering
        \includegraphics[width=\linewidth]{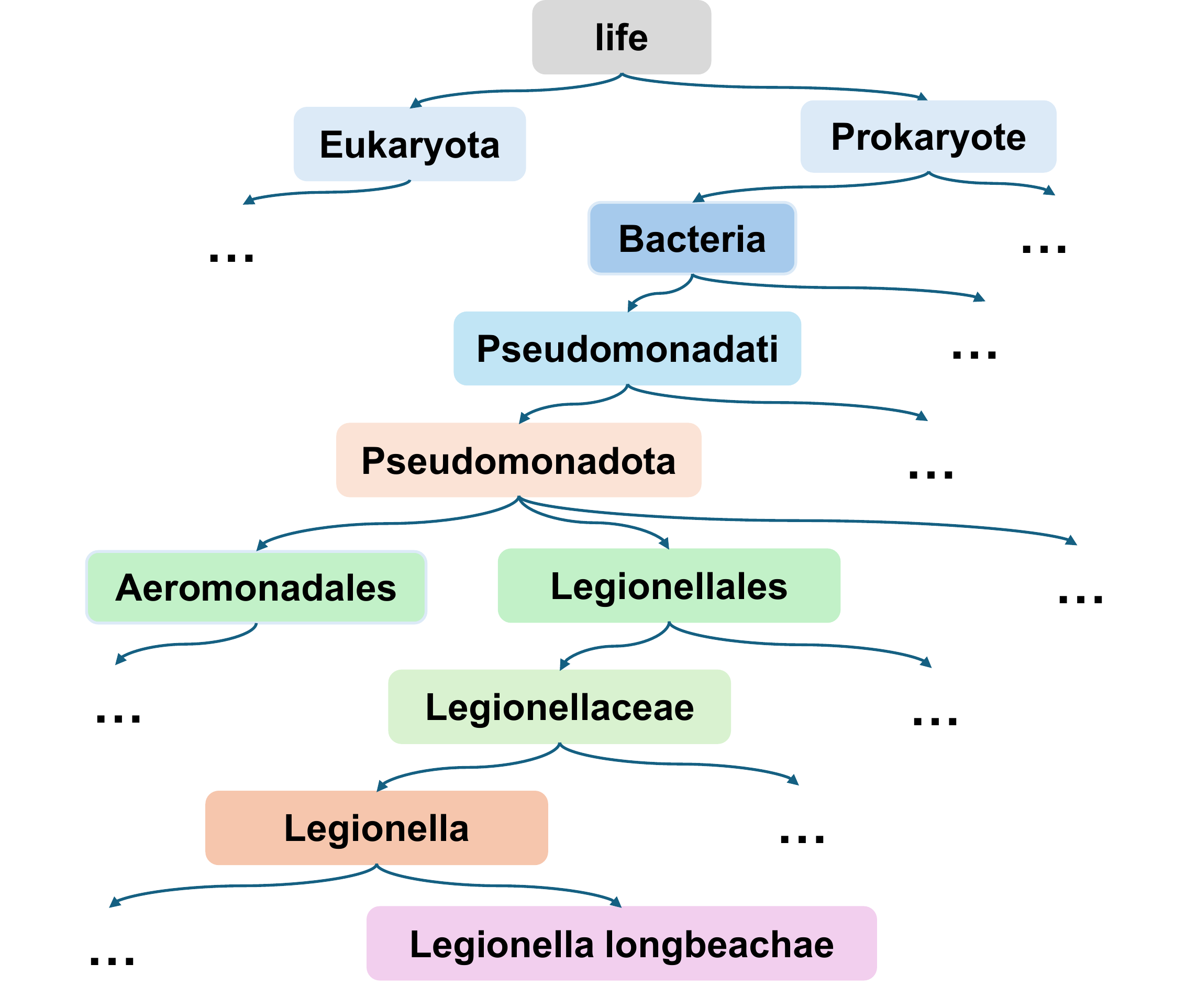}
        \smallskip

        {\small\textbf{(a)} Local taxonomic neighborhood of \textit{Legionella longbeachae}.}
    \end{minipage}
    \hfill
    \begin{minipage}[t]{0.49\textwidth}
        \centering
        \includegraphics[width=\linewidth]{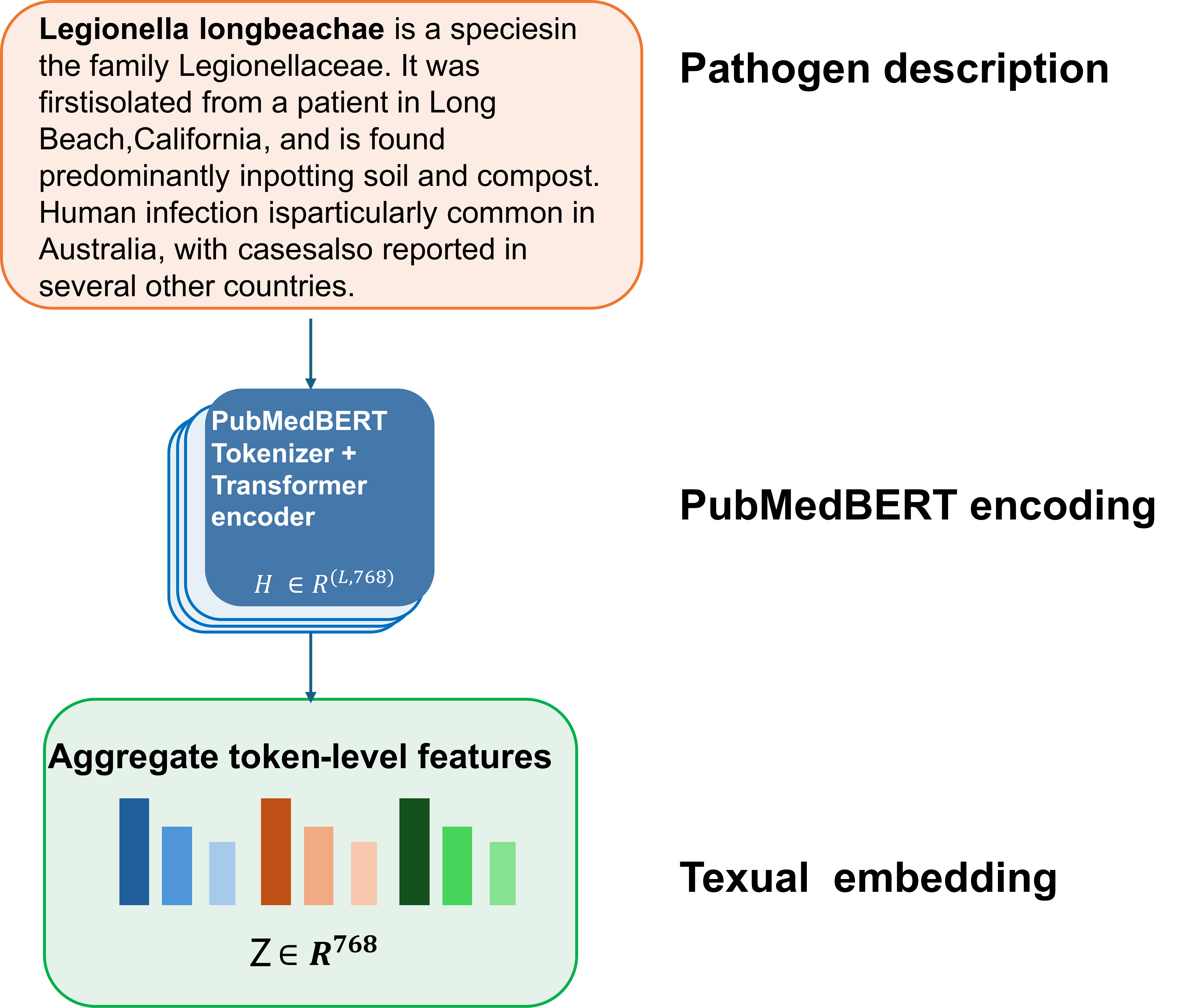}
        \smallskip

        {\small\textbf{(b)} PubMedBERT-based construction of pathogen text embeddings.}
    \end{minipage}

    \caption{Structural and semantic pathogen representations in PathoMIC. Panel (a) illustrates a local neighborhood in the pathogen taxonomy graph. Panel (b) shows the pipeline used to encode a pathogen description into a 768-dimensional text embedding with PubMedBERT and mean pooling.}
    \Description{Two panels show the local taxonomy surrounding Legionella longbeachae and the pipeline that converts a pathogen description into a 768-dimensional text embedding using PubMedBERT and mean pooling.}
    \label{fig:pathogen-representations}
\end{figure*}

Multiple public databases have been developed to accommodate the growing number of experimentally measured antimicrobial peptide (AMP) records, including APD6, DBAASP, DRAMP, CAMP3, and
YADAMP~\cite{APD,DBAASP,DRAMP,CAMPR3,YADAMP}. However, these resources are not
organized under a unified schema: MIC values are reported in heterogeneous
units, target organisms are named inconsistently, and some records appear only
in free-text descriptions rather than structured tables. These inconsistencies
make it difficult to directly compare measurements across sources or construct
reliable peptide--pathogen prediction benchmarks. A unified curation pipeline is
therefore necessary to standardize the data and improve its usability for
cross-species MIC modeling.

We constructed PathoMIC by consolidating quantitative AMP MIC measurements from
the five public resources listed above. We selected these databases because
they contain experimentally validated records, are widely used in AMP research,
and collectively cover a broad range of pathogen species. We further
incorporated the recently curated HMAMP and GRAMPA collections~\cite{2025HMAMP,grampa}, whose tabular organization facilitates structured data
integration. To expand the scale and taxonomic diversity of the dataset, we
also retrieved relevant entries from UniProt~\cite{UniProt} using the keyword
\texttt{ANTIMICROBIAL}. The dataset snapshot includes records available from
all sources as of January~26, 2026.

We developed source-specific parsers and web crawlers to extract peptide
sequences, reported MIC values, and target-organism names and then mapped the
heterogeneous source fields to a unified schema. This standardization enables
records from different databases to be compared and processed consistently. We
resolved target organisms to standardized scientific species names through
string matching and name normalization, thereby reducing duplicate or
ambiguous pathogen identities.

MIC measurements reported in heterogeneous concentration units were converted
to micromolar ($\mu\mathrm{M}$), which provides a common scale across sources.
We represent antimicrobial potency as
\[
y=\log_{10}\left(\frac{\mathrm{MIC}}{1\,\mu\mathrm{M}}\right),
\]
which compresses the wide dynamic range of MIC values while preserving their
ordering. After merging all sources, we treated each normalized
peptide--pathogen pair as a unique entity. Duplicate observations can arise from
overlapping databases or repeated assays conducted under different
experimental conditions. We aggregated such observations using the median MIC,
which is more robust to extreme values than the mean. As a result, each retained
peptide--pathogen pair is associated with one harmonized MIC target, reducing
redundancy and improving the stability of downstream evaluation, as shown in Figure~\ref{fig:dataset-curation-pipeline}(a).

The finalized PathoMIC dataset contains 74,751 peptide--pathogen MIC records,
11,125 distinct peptide sequences, and 424 pathogen species spanning three
major groups: Gram-negative bacteria, Gram-positive bacteria, and fungi. Detailed statistics on the composition of each pathogen group are presented in Figure~\ref{fig:dataset-composition}(b).
Peptide lengths range from 1 to 190 amino acids, with a median of 18. The
transformed MIC values range from $-4.6842$ to $4.8368$. 
Figure~\ref{fig:dataset-composition}(c) shows a pronounced long-tailed
distribution across pathogen species. A small number of extensively studied
species account for a large fraction of the observations, whereas many species
have only a few measurements. Although this imbalance reflects the current
state of experimental AMP research, it also creates a realistic setting for
evaluating whether models can transfer knowledge from data-rich pathogens to
low-resource or previously unseen species.

\subsection{Pathogen Knowledge Augmentation}

For each pathogen species, we queried the NCBI Taxonomy database~\cite{NCBI} to
obtain its TaxID and taxonomic lineage. We projected each lineage onto eight
canonical ranks---domain, kingdom, phylum, class, order, family, genus, and
species---and merged ancestors shared across species. Connecting each retained
parent taxon to its child produced a directed acyclic taxonomy graph in which
nearby pathogen nodes share increasingly specific ancestors.  Figure~\ref{fig:pathogen-representations}(a)
illustrates a local neighborhood centered on \textit{Legionella longbeachae}, 
including its lineage and nearby species. We construct the taxonomy graph from the 366 pathogen names included in the downstream experiments. As shown in Figure~\ref{fig:complete-taxonomy}, the resulting graph contains 709 nodes and 707 directed edges. Although all 366 pathogen names are retained in the pathogen-to-node mapping, they correspond to 364 distinct species-rank leaf nodes. Specifically, \textit{Salmonella enterica}, \textit{Salmonella enteritidis}, and \textit{Salmonella typhimurium} are mapped to the same species-level node, \textit{Salmonella enterica}. The latter two taxa are classified as serovars below the species rank. Because we retain only eight canonical taxonomic ranks, their serovar-level distinctions are removed and their lineages converge at the same species-level ancestor. The graph therefore comprises 2 domain-level, 4 kingdom-level, 11 phylum-level, 29 class-level, 58 order-level, 90 family-level, 151 genus-level, and 364 species-level nodes. 

The taxonomy graph provides an explicit representation of hierarchical biological relationships among pathogens. Unlike independent species identifiers, which treat all pathogens as equally unrelated categories, the graph allows closely related species to share information through their common taxonomic ancestors. This structure is particularly valuable for pathogens with limited MIC measurements, as their representations can be informed by data-rich relatives at the genus, family, or higher taxonomic levels. It therefore provides a biologically grounded inductive bias for sharing information across taxonomically related pathogens, which may be particularly useful when target-species MIC measurements are scarce or unavailable.

To complement this discrete taxonomic structure with species-level semantic
information, we collected a textual description of each pathogen from its
public Wikipedia page. As illustrated in
Figure~\ref{fig:pathogen-representations}(b), each description is tokenized and
encoded with PubMedBERT~\cite{pubmedbert}, producing contextual token
representations in $\mathbb{R}^{L\times768}$ for a description containing $L$
tokens. Mean pooling over the token dimension, followed by L2 normalization,
produces a fixed 768-dimensional vector
$\mathbf{z}_{\mathrm{text}}\in\mathbb{R}^{768}$ for each pathogen. Together, the taxonomy graph and textual embeddings provide complementary sources of pathogen knowledge. The former captures structured relationships shared across taxonomic levels, while the latter preserves fine-grained semantic attributes specific to individual pathogens. Incorporating both representations enables a model to move beyond memorizing pathogen identities and instead learn biologically meaningful similarities that support cross-species MIC prediction.

%% file: sections/Methods.tex
\section{Methods}

\providecommand{\Raw}{\textsc{Raw}}
\providecommand{\SP}{\textsc{SP}}
\providecommand{\Shuffle}{\textsc{Shuffle}}
\providecommand{\GNN}{\textsc{GNN}}

\subsection{Peptide Encoder}

We model MIC prediction as a regression problem over peptide--pathogen pairs.
Given a peptide sequence, we use the pretrained protein language model
ESM2~\cite{ESM2} to obtain residue-level representations. Each sequence is
right-padded with the ambiguous-residue token \texttt{X} to a fixed maximum
length before tokenization. This preprocessing scheme is shared by all model
variants to ensure a controlled comparison.
Let
$\mathbf{H}=[\mathbf{h}_1,\ldots,\mathbf{h}_T]
\in\mathbb{R}^{T\times d_{\mathrm{esm}}}$
denote the final-layer hidden states produced by ESM2. We obtain a fixed-length
peptide representation through mean pooling:
\begin{equation}
  \mathbf{z}_{\mathrm{pep}}
  =\frac{1}{T}\sum_{t=1}^{T}\mathbf{h}_t
  \in\mathbb{R}^{640}.
  \label{eq:meanpool}
\end{equation}

We use a three-layer MLP regression head, consisting of two hidden layers with widths 256 and 64,
each followed by ReLU activation and one-dimensional batch normalization, and a
final linear projection to a scalar. The output layer uses neither an activation function nor normalization because
the log-transformed MIC target is unbounded and may take negative values.

\subsection{Pathogen-Aware Model Variants}

To evaluate the value of the pathogen descriptions and taxonomy graph provided
by PathoMIC, we augment the peptide encoder with several pathogen-information
pathways. The variants differ only in how pathogen information is represented
and integrated.

  \textbf{Sequence-only reference: Raw} receives no pathogen information. 
  The pooled peptide representation is passed directly to the regression head. 
  Because the model is agnostic to the target organism, it predicts the same MIC
for a peptide regardless of the pathogen with which the peptide is paired.

  \textbf{Pathogen sepecies description adapter: SP} introduces a species-adapter pathway for pathogen descriptions. 
  A learnable nonlinear species adapter maps the fixed description embedding to a
128-dimensional pathogen representation:
\begin{equation}
  \mathbf{z}_{\mathrm{sp}}
  =
  \mathrm{GELU}\!\left(
    \mathbf{W}_2
    \mathrm{Drop}_{0.1}\!\left[
      \mathrm{GELU}\!\left(
        \mathrm{LN}(\mathbf{W}_1\mathbf{s}_{\mathrm{txt}}+\mathbf{b}_1)
      \right)
    \right]
    +\mathbf{b}_2
  \right)
  \in\mathbb{R}^{128},
  \label{eq:adapter}
\end{equation}
where
$\mathbf{W}_1\in\mathbb{R}^{128\times768}$ and
$\mathbf{W}_2\in\mathbb{R}^{128\times128}$. The adapter output is concatenated
with the peptide representation:
\begin{equation}
  \mathbf{u}
  =
  [\mathbf{z}_{\mathrm{pep}};\mathbf{z}_{\mathrm{sp}}]
  ,
  \label{eq:concat-sp}
\end{equation}
and passed to the shared regression head.

  \textbf{Negative control for peptide--pathogen correspondence: Shuffle} is architecturally identical to SP. Before training, we replace
the pathogen column of the training split with a uniformly random permutation
drawn once using a fixed seed. Peptide sequences and MIC targets remain
unchanged. Because the operation is a permutation, it preserves the marginal
frequency of each pathogen description while disrupting the true
peptide--pathogen correspondence. Validation and test pairs retain their
correct pathogen labels. This control therefore preserves the architecture,
input dimensionality, description embeddings, and label distribution while
removing meaningful pathogen conditioning.

  \textbf{Taxonomy-aware pathogen encoder: GNN} propagates pathogen information over the taxonomy graph so that
taxonomically related organisms can share statistical strength. 
Every graph node is initialized with a 768-dimensional PubMedBERT embedding.
Species nodes reuse the description embeddings used by \SP{} when available;
other nodes are encoded from rank-qualified taxon names, such as
``genus \textit{Staphylococcus}.'' These initial features are frozen during
training so that only the graph encoder learns to transform and propagate them.
We use a two-layer graph-convolution network~\cite{GCN} (GCN) and apply the nonlinearity and dropout only between the two GCN 
layers.

For a pathogen $p$, we retrieve the learned embeddings of its species, genus,
and family nodes and combine them through a learnable hierarchical projection:
\begin{equation}
  \mathbf{z}_{\mathrm{sp}}
  =
  \mathbf{W}_{\mathrm{hier}}
  [
    \mathbf{e}_{\mathrm{species}(p)};
    \mathbf{e}_{\mathrm{genus}(p)};
    \mathbf{e}_{\mathrm{family}(p)}
  ]
  +\mathbf{b}_{\mathrm{hier}},
  \label{eq:hier}
\end{equation}
If NCBI does not assign one of these ranks, we substitute a rank-specific
learnable missing-rank vector initialized from
$\mathcal{N}(0,0.02^2)$. The pathogen representation is then concatenated with
the peptide representation:
\begin{equation}
  \mathbf{u}
  =
  [\mathbf{z}_{\mathrm{pep}};\mathbf{z}_{\mathrm{sp}}], 
  \label{eq:concat-gnn}
\end{equation}
and passed to the shared regression head.

  \textbf{Semantics-free node-feature control: GNN-RI} retains the taxonomy graph, two-layer GCN, hierarchical readout, and
frozen-feature setting of \GNN{}, but replaces the semantic node features with
fixed random vectors. Specifically, before training, we sample
\begin{equation}
  \widetilde{\mathbf{X}}^{(0)}_{ij}
  \sim\mathcal{N}(0,\sigma^2),
  \qquad
  \sigma=\mathrm{std}(\mathbf{X}^{(0)})\approx0.372,
  \label{eq:randinit}
\end{equation}
so that the random features approximately match the scale of the PubMedBERT
features. The random matrix is sampled once using a fixed seed and remains
frozen throughout training. Consequently, each node receives a stable but
semantically arbitrary identifier, while the model can still exploit the graph
topology and hierarchical readout.
Because \GNN{} and \GNN-RI{} share the same topology, dimensions, and training
procedure, their performance difference isolates the contribution of semantic
node initialization.

%% file: sections/Evaluation_Framework.tex
\section{Evaluation Framework}

PathoMIC exhibits a strongly long-tailed distribution across target pathogens,
making conventional random record-level evaluation insufficient for assessing
cross-species generalization. We therefore establish two complementary protocols. Few-shot evaluation measures
prediction for species with limited training supervision, whereas zero-shot
evaluation evaluates prediction for species that are entirely absent from the
training and validation sets. 

For all experiments, we first retain only records whose peptide
sequences contain at most 32 amino acids. Ribosomally synthesized AMPs are
typically shorter than 50 residues~\cite{OliveiraJunior2025AMP}. Moreover,
computational AMP design studies have explicitly favored shorter candidates to
reduce synthesis cost or improve ease of synthesis~\cite{Vishnepolsky2019DeNovo,Boone2021GAAMP}.
In PathoMIC,  75\% of peptides contain at most 24 amino acids; thus, the
32-residue threshold retains most of the dataset while focusing the benchmark
on practically synthesizable candidates. After resolving repeated measurements by median aggregation, the 74,751 normalized MIC observations yield 55,253 unique peptide--pathogen records. Applying the peptide-length criterion further retains 48,429 of these 55,253 records (87.65\%), while pathogen coverage decreases from 424 to 366 species. 

\subsection{Cross-Species Evaluation Protocols}

\textbf{Zero-shot evaluation.}
Within the filtered dataset, we identify all pathogen species with fewer than
10 MIC records and assign all of their peptide--pathogen pairs to the zero-shot
test set. No record associated with these species appears in the training or
validation set. This species-level holdout prevents the model from observing
labeled MIC measurements for the target species during optimization and
directly evaluates transfer to unseen pathogens. \textbf{Few-shot evaluation.}
For each remaining pathogen species, we partition its MIC records into
training, validation, and test sets using an 8:1:1 ratio. Splitting within each
species maintains broad pathogen coverage across the three partitions while
preserving the naturally long-tailed distribution of record counts. We define
a species as a few-shot target if fewer than 20 of its MIC records are assigned
to the training set. Performance on the corresponding held-out test pairs
measures prediction under limited target-species supervision.

Together, these protocols capture two target-species supervision regimes. The few-shot setting evaluates prediction with scarce
target-species supervision, whereas the zero-shot setting evaluates transfer
without any labeled MIC measurements for the target species.

\subsection{Evaluation Metrics}

We evaluate predictions of the $\log_{10}$-transformed MIC values using three
complementary metrics: mean squared error (MSE), Spearman's rank correlation
coefficient $\rho$, and Kendall's rank correlation coefficient $\tau$. For
$N$ evaluated peptide--pathogen pairs with targets $y_i$ and predictions
$\hat{y}_i$, MSE is defined as
\begin{equation}
    \mathrm{MSE}
    = \frac{1}{N}\sum_{i=1}^{N}\left(y_i-\hat{y}_i\right)^2.
\end{equation}
MSE is also used as the training objective and measures absolute prediction
error; lower values indicate better performance.

Spearman's $\rho$ measures the correlation between the ranks of predicted and
observed MIC values, whereas Kendall's $\tau$ measures pairwise rank
concordance. Both coefficients range from $-1$ to $1$, with larger values
indicating better preservation of relative antimicrobial potency across
peptide--pathogen pairs. Thus, MSE evaluates absolute prediction accuracy,
whereas Spearman's $\rho$ and Kendall's $\tau$ evaluate ranking quality.
Unless otherwise stated, we report the mean and standard deviation of each
metric over three independent runs.

%% file: sections/Experiments.tex
\section{Experiments}
In this section, we evaluate a range of MIC prediction models to characterize both the potential and the remaining challenges of cross-species MIC prediction in few-shot and zero-shot settings. We further analyze the effects of the pathogen descriptions and taxonomic relationships provided by PathoMIC.
\begin{table*}[htpb]
  \centering
  \caption{Sequence-only MIC prediction results under few-shot and zero-shot settings. All models receive only peptide sequences and no pathogen information. Best results are shown in bold, and second-best results are underlined.}
  \label{tab:sequence-only-mic}
  \small
  \setlength{\tabcolsep}{4.2pt}
  \begin{tabular}{llccc|ccc}
    \toprule
    & & \multicolumn{3}{c|}{Few-shot} & \multicolumn{3}{c}{Zero-shot} \\
    \cmidrule(lr){3-5}\cmidrule(lr){6-8}
    Model & Backbone
      & MSE $\downarrow$ & Spearman's $\rho$ $\uparrow$ & Kendall's $\tau$ $\uparrow$
      & MSE $\downarrow$ & Spearman's $\rho$ $\uparrow$ & Kendall's $\tau$ $\uparrow$ \\
    \midrule
    \rowcolor{gray!15}
    \multicolumn{8}{c}{\textit{General deep learning models}} \\
    AMP-Designer & GPT-2 & 0.2687 $\pm$ 0.0079 & 0.7134 $\pm$ 0.0091 & 0.5327 $\pm$ 0.0093 & 0.4658 $\pm$ 0.0219 & 0.5826 $\pm$ 0.0207 & 0.4182 $\pm$ 0.0186 \\
    GRAMPA       & CNN   & 0.2663 $\pm$ 0.0026 & 0.7145 $\pm$ 0.0020 & 0.5332 $\pm$ 0.0017 & 0.4594 $\pm$ 0.0080 & 0.5789 $\pm$ 0.0081 & 0.4154 $\pm$ 0.0071 \\
    LSTM-Seq     & LSTM  & 0.2598 $\pm$ 0.0036 & 0.7234 $\pm$ 0.0051 & \textbf{0.5687 $\pm$ 0.0059} & 0.4483 $\pm$ 0.0050 & 0.6008 $\pm$ 0.0012 & 0.4364 $\pm$ 0.0016 \\
    \addlinespace[2pt]
    \rowcolor{gray!15}
    \multicolumn{8}{c}{\textit{Pretrained protein language models}} \\
    ESM2-8M   & PLM & 0.2493 $\pm$ 0.0016 & 0.7327 $\pm$ 0.0016 & 0.5532 $\pm$ 0.0029 & 0.4338 $\pm$ 0.0039 & 0.6021 $\pm$ 0.0074 & 0.4378 $\pm$ 0.0057 \\
    ESM2-35M  & PLM & \underline{0.2484 $\pm$ 0.0021} & \textbf{0.7345 $\pm$ 0.0032} & \underline{0.5553 $\pm$ 0.0039} & \underline{0.4249 $\pm$ 0.0085} & \textbf{0.6128 $\pm$ 0.0083} & \textbf{0.4475 $\pm$ 0.0080} \\
    ESM2-150M & PLM & \textbf{0.2479 $\pm$ 0.0021} & \underline{0.7329 $\pm$ 0.0015} & 0.5536 $\pm$ 0.0019 & \textbf{0.4220 $\pm$ 0.0041} & \underline{0.6115 $\pm$ 0.0039} & \underline{0.4446 $\pm$ 0.0050} \\
    \bottomrule
  \end{tabular}
\end{table*}

\subsection{Sequence-Only MIC Prediction}
We first evaluate whether peptide sequences alone can support MIC prediction when no pathogen information is provided. Under this sequence-only setting, we consider two categories of representative models. The first category comprises general deep learning approaches, including the LSTM-based LSTM-Seq~\cite{LSTM_Seq, lstm}, the CNN-based GRAMPA~\cite{grampa,CNN}, and AMP-Designer~\cite{AMP-Designer}, which uses a GPT-2~\cite{GPT2} backbone. The second category comprises pretrained protein language models. Specifically, we evaluate ESM2~\cite{ESM2} variants with 8M, 35M, and 150M parameters and attach a three-layer MLP regressor to each backbone. For every model, we report the mean and standard deviation of MSE, Spearman's $\rho$, and Kendall's $\tau$ over three independent runs in both few-shot and zero-shot settings, as shown in Table~\ref{tab:sequence-only-mic}.

The results yield three main observations. First, pretrained protein language models provide the most consistent improvements in absolute MIC estimation under both evaluation settings. 
The ESM2 variants achieve the three lowest MSE values in both few-shot and zero-shot evaluation.
Except for that LSTM-Seq performs best on few-shot Kendall's $\tau$, the ESM2 variants generally 
outperform the conventional deep learning baselines on the two rank-correlation metrics.  
These results suggest that pretrained protein representations capture sequence features that are useful 
for MIC prediction, which may be due to their exposure to large peptide corpora during pretraining. 

Second, within the ESM2 family, increasing the model size from 8M to 150M parameters monotonically 
reduces MSE in both settings. ESM2-150M achieves the lowest MSE, improving over the strongest general 
deep learning baseline, LSTM-Seq, by $4.6\%$ in the few-shot setting and $5.9\%$ in the zero-shot setting. 
This trend suggests that larger models may improve the calibration of predicted MIC values.

Third, every model performs worse under zero-shot evaluation than under few-shot evaluation. 
Across models, MSE increases by approximately $0.17$--$0.20$, while Spearman's $\rho$ and Kendall's $\tau$ 
decrease by approximately $0.12$--$0.14$. This consistent degradation indicates that peptide sequences 
alone do not fully account for distribution shifts across pathogen species, motivating the explicit 
incorporation of pathogen information in the following experiments.

\subsection{MIC Prediction with Pathogen Information}
\begin{table*}[htpb]
  \centering
  \caption{Effect of pathogen information on ESM2-150M under few-shot and zero-shot cross-species MIC prediction. Best results are shown in bold, and second-best results are underlined.}
  \label{tab:pathogen-information}
  \small
  \setlength{\tabcolsep}{4.2pt}
  \begin{tabular}{lccc|ccc}
    \toprule
    & \multicolumn{3}{c|}{Few-shot} & \multicolumn{3}{c}{Zero-shot} \\
    \cmidrule(lr){2-4}\cmidrule(lr){5-7}
    Model
      & MSE $\downarrow$ & Spearman's $\rho$ $\uparrow$ & Kendall's $\tau$ $\uparrow$
      & MSE $\downarrow$ & Spearman's $\rho$ $\uparrow$ & Kendall's $\tau$ $\uparrow$ \\
    \midrule
    ESM2-150M-Raw
      & 0.2479 $\pm$ 0.0021 & 0.7329 $\pm$ 0.0015 & 0.5536 $\pm$ 0.0019
      & \underline{0.4220 $\pm$ 0.0041} & 0.6115 $\pm$ 0.0039 & 0.4446 $\pm$ 0.0050 \\
    ESM2-150M-Shuffle
      & 0.2574 $\pm$ 0.0060 & 0.7268 $\pm$ 0.0058 & 0.5470 $\pm$ 0.0062
      & 0.4381 $\pm$ 0.0033 & 0.5977 $\pm$ 0.0077 & 0.4345 $\pm$ 0.0060 \\
    ESM2-150M-GNN
      & \underline{0.1675 $\pm$ 0.0060} & 0.8239 $\pm$ 0.0031 & 0.6480 $\pm$ 0.0032
      & \textbf{0.4118 $\pm$ 0.0148} & \textbf{0.6305 $\pm$ 0.0126} & \textbf{0.4636 $\pm$ 0.0091} \\
    ESM2-150M-GNN-RI
      & 0.1677 $\pm$ 0.0060 & \underline{0.8360 $\pm$ 0.0020} & \underline{0.6616 $\pm$ 0.0027}
      & 0.4249 $\pm$ 0.0117 & \underline{0.6233 $\pm$ 0.0066} & \underline{0.4595 $\pm$ 0.0060} \\
    ESM2-150M-SP
      & \textbf{0.1528 $\pm$ 0.0039} & \textbf{0.8474 $\pm$ 0.0034} & \textbf{0.6784 $\pm$ 0.0050}
      & 0.4442 $\pm$ 0.0048 & 0.6227 $\pm$ 0.0064 & 0.4575 $\pm$ 0.0054 \\
    \bottomrule
  \end{tabular}
\end{table*}

Having identified ESM2-150M as the strongest sequence-only regressor, we use it as the peptide encoder in all subsequent experiments and investigate whether pathogen descriptions and taxonomic relationships improve cross-species MIC prediction. 
We evaluate all variants under both few-shot and zero-shot settings and report the mean and standard deviation over three independent runs in Table~\ref{tab:pathogen-information}. The results yield four main observations.

First, correctly matched pathogen descriptions provide substantial gains when target-species supervision is available. In the few-shot setting, ESM2-150M-SP reduces MSE from $0.2479$ to $0.1528$, corresponding to a relative reduction of $38.4\%$, while increasing Spearman's $\rho$ by $0.1145$ and Kendall's $\tau$ by $0.1248$. It achieves the best result on all three few-shot metrics, indicating that pathogen descriptions provide complementary information beyond peptide sequences.

Second, randomly mismatched pathogen descriptions consistently harm performance. ESM2-150M-Shuffle performs worse than ESM2-150M-Raw on all six metrics. Because the two variants differ primarily in whether the added pathogen description is correctly matched, this control indicates that the gains from pathogen-aware modeling do not arise merely from additional parameters or a larger input representation.

Third, among the pathogen-aware variants, taxonomic message passing performs most favorably in the zero-shot setting. ESM2-150M-GNN achieves the best zero-shot MSE ($0.4118$), Spearman's $\rho$ ($0.6305$), and Kendall's $\tau$ ($0.4636$). ESM2-150M-GNN-RI remains competitive and ranks second on both zero-shot correlation metrics, suggesting that the taxonomy topology may provide a useful relational signal even without semantic node features. In the few-shot setting, however, GNN-RI slightly exceeds GNN on both rank-correlation metrics, whereas their MSE values are nearly identical. These results suggest that semantic descriptions and taxonomic structure play complementary roles across the two evaluation regimes.

Finally, the magnitude of improvement differs substantially between few-shot and zero-shot evaluation. Relative to ESM2-150M-Raw, the best few-shot model reduces MSE from $0.2479$ to $0.1528$ ($38.4\%$), whereas the best zero-shot model reduces MSE from $0.4220$ to $0.4118$ ($2.4\%$). Thus, pathogen information is highly effective when limited target-species supervision is available but provides only modest gains when the target species is entirely unseen. This remaining gap highlights the difficulty of zero-shot cross-species MIC prediction.

\subsection{Long-tail Analysis across Pathogens}
\begin{figure*}[htbp]
  \centering
  \begin{minipage}[t]{0.326\textwidth}
    \vspace{0pt}
    \centering
    \includegraphics[width=\linewidth]{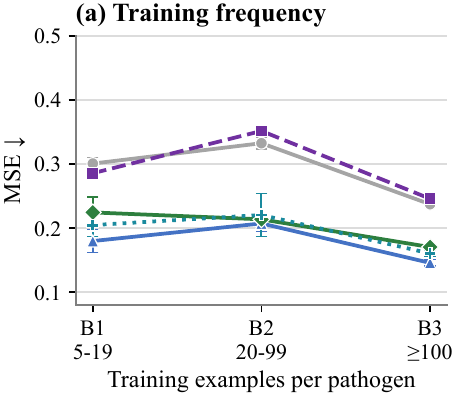}
  \end{minipage}\hfill
  \begin{minipage}[t]{0.326\textwidth}
    \vspace{0pt}
    \centering
    \includegraphics[width=\linewidth]{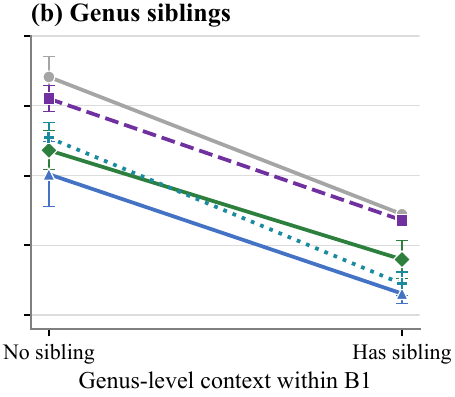}
  \end{minipage}\hfill
  \begin{minipage}[t]{0.326\textwidth}
    \vspace{0pt}
    \centering
    \includegraphics[width=\linewidth]{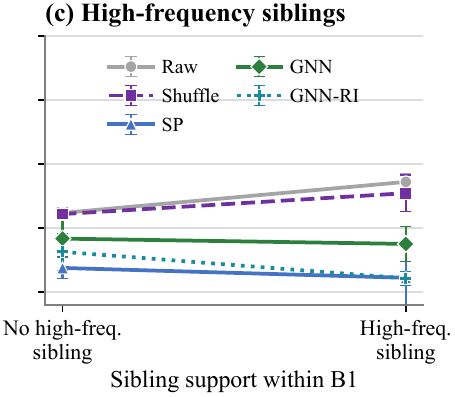}
  \end{minipage}
  \Description{Three point-and-line plots compare five ESM2-150M variants in
  the few-shot long-tail analysis. Panel (a) groups test pathogens by training
  frequency. Panel (b) compares B1 pathogens with and without another training
  species from the same genus. Panel (c) divides B1 pathogens with genus-level
  siblings according to whether at least one sibling belongs to B3.}
  \caption{Few-shot MIC prediction under long-tailed pathogen distributions.
  (a) Results for B1, B2, and B3. (b) Results for B1 pathogens with and without
  a genus-level sibling in the training set. (c) Results for B1 pathogens with
  siblings, grouped by whether a high-resource sibling is available.}
  \label{fig:long-tail}
\end{figure*}

To evaluate robustness under the long-tailed pathogen distribution, we first
assign each test pathogen to a bucket according to its number of training
examples. We define three bunckets: B1 contains low-resource pathogens with 5--19 training examples, B2
contains pathogens with 20--99 examples, and B3 contains high-resource
pathogens with at least 100 examples. For each bucket, we pool all test
peptide--pathogen pairs associated with its pathogens and average their squared
prediction errors. Tables~\ref{tab:long-tail-buckets} summarizes the number of pathogens and test pairs in each bucket.
We further partition B1 by genus-level taxonomic context. A pathogen is
categorized as \emph{no sibling} if no other training species from the same
genus is represented and as \emph{has sibling} otherwise. Among pathogens with
siblings, we additionally distinguish whether at least one sibling belongs to
the high-resource B3 group and categorize pathogens accordingly as \emph{high-resource sibling} or \emph{no high-resource sibling}. 
Table~\ref{tab:b1-genus-composition} summarizes the genus-level composition of B1.

\begin{table}[htbp]
  \centering
  \caption{Pathogen-frequency groups based on the number of training examples.}
  \label{tab:long-tail-buckets}
  \small
  \setlength{\tabcolsep}{12.0pt}
  \begin{tabular}{lrr}
    \toprule
    Bucket & \# Pathogens & \# Test pairs \\
    \midrule
    B1: $[5,20)$       & 66 (37.3\%) & 158 (3.2\%) \\
    B2: $[20,100)$     & 72 (40.7\%) & 486 (9.8\%) \\
    B3: $[100,\infty)$ & 39 (22.0\%) & 4,299 (87.0\%) \\
    \midrule
    Total              & 177         & 4,943 \\
    \bottomrule
  \end{tabular}
\end{table}

\begin{table}[thbp]
  \centering
  \caption{Genus-level sibling composition of B1 pathogens, further grouped by the availability of high-resource siblings.}
  \label{tab:b1-genus-composition}
  \small
  \setlength{\tabcolsep}{12.0pt}
  \begin{tabular}{lrr}
    \toprule
    Group & \# Pathogens & \# Test pairs \\
    \midrule
    No sibling                     & 18 (27.3\%) & 45 (28.5\%) \\
    Has sibling                    & 48 (72.7\%) & 113 (71.5\%) \\
    \quad High-resource sibling    & 21 (43.8\%) & 49 (43.4\%) \\
    \quad No high-resource sibling & 27 (56.3\%) & 64 (56.6\%) \\
    \bottomrule
  \end{tabular}
\end{table}

Figure~\ref{fig:long-tail}(a) compares performance across pathogen-frequency
groups. All variants perform best on B3, indicating that greater
pathogen-specific supervision is associated with lower prediction error.
ESM2-150M-SP achieves the lowest MSE in every bucket, showing that matched
pathogen descriptions remain useful across resource levels.
Figures~\ref{fig:long-tail}(b) and~\ref{fig:long-tail}(c) examine taxonomic
context within B1. Every model performs better on pathogens with genus-level
siblings than on those without siblings. 
The high-resource-sibling comparison provides stronger evidence of transfer.
SP, GNN, and GNN-RI improve when a B3 sibling is available, whereas Raw and
Shuffle exhibit the opposite trend. GNN-RI performs best in the
high-resource-sibling group, while SP performs best when no B3 sibling is
available. This contrast suggests that semantic pathogen descriptions provide
broad benefits for low-resource prediction, whereas taxonomic message passing
is particularly effective when closely related, well-represented species are
available.

%% file: sections/Conclusion_and_Future_Work.tex
\section{Conclusion and Future Work}
This paper introduces PathoMIC as the largest and most pathogen-diverse unified
resource for quantitative AMP activity modeling across pathogens. PathoMIC enables a standardized cross-species benchmark
that exposes the challenges of generalization under long-tailed supervision.
Our results show that pathogen knowledge provides substantial predictive benefits when limited target-species supervision is available. For entirely unseen pathogens, taxonomy-aware modeling yields only modest improvements, indicating that robust zero-shot transfer remains an open challenge. Together, PathoMIC and the accompanying benchmark provide a
foundation for developing and evaluating pathogen-aware AMP predictors.

We would like to acknowledge the limitations of the current study. The best pathogen-aware variant improves the zero-shot MSE by only 2.4\% relative to the sequence-only baseline, and not all pathogen-aware variants consistently improve zero-shot performance. Robust generalization 
to entirely unseen pathogens remains unresolved. 
Despite this limitation, PathoMIC could serve as an important resource for
developing and evaluating computational screening systems for emerging or rare
pathogens and for prioritizing in silico designed AMPs before experimental
validation~\cite{Vishnepolsky2019DeNovo}. A central direction is to improve zero-shot transfer through
transfer learning, domain adaptation, meta-learning, and uncertainty-aware
prediction, while exploiting external taxonomic, genomic, and textual knowledge
without relying on target-species MIC labels~\cite{BERTAmPEP60}. Another direction is to integrate
additional AMP resources, including those with experimentally annotated toxicity
and stability data~\cite{DBAASP,DRAMP}, and move from MIC-only regression toward multi-task,
multi-property prediction. Jointly modeling antimicrobial potency with
hemolytic activity such as HC50, mammalian-cell cytotoxicity, stability,
solubility, and related developability properties would enable models to rank
candidates by both efficacy and safety~\cite{2025HMAMP}. Combined with prospective wet-lab
validation, these extensions could make PathoMIC a foundation for more
reliable, pathogen-specific, and experimentally efficient AMP discovery.